\documentclass[prc,onecolumn,tightenlines,12pt]{revtex4}
\usepackage{graphicx}
\usepackage{dcolumn}
\usepackage{amsmath}
\usepackage{epsfig}
\usepackage{SIunits}
\newcommand{\be}{\begin{equation}}
\newcommand{\ee}{\end{equation}}
\newcommand{\bea}{\begin{eqnarray}}
\newcommand{\eea}{\end{eqnarray}}

\newcommand{\rref}[1]{Ref.~\cite{#1}}
\newcommand{\fref}[1]{Fig.~\ref{#1}}
\newcommand{\eref}[1]{Eq.~(\ref{#1})}
\newcommand{\erefs}[2]{Eqs.~(\ref{#1}-\ref{#2})}
\newcommand{\sref}[1]{section~\ref{#1}}
\newcommand{\aref}[1]{appendix~\ref{#1}}
\newcommand{\tref}[1]{table~\ref{#1}}

\begin{document}

\author{M.~S.~Hussein$^{\dag 1,2}$, W.~Li$^{1}$ and S.~W\"uster$^{1}$}
\affiliation{$^{1}$Max-Planck-Institut f\"ur Physik komplexer Systeme\\
N\"othnitzer Stra$\beta$e 38, D-01187 Dresden, Germany \\
$^{2}$Instituto de F\'{i}sica, Universidade de S\~{a}o Paulo\\
C.P. 66318, 05315-970 S\~{a}o Paulo, S.P., Brazil}
\title{Canonical quantum potential scattering theory
\thanks{Supported in part by the CNPq and FAPESP (Brazil).\\
        $^{\dag}$Martin Gutzwiller Fellow, 2007/2008.}}

\begin{abstract}
A new formulation of potential scattering in quantum mechanics is developed using a close structural 
analogy between partial waves and the classical dynamics of many non-interacting fields. Using a canonical formalism we find non-linear first-order differential equations for the low energy scattering parameters like scattering length and effective range.
They significantly simplify typical calculations, as we illustrate for atom-atom and neutron-nucleus scattering systems. A generalization to charged particle scattering is also possible.
\end{abstract}


\maketitle

\section{Introduction}

Low-energy scattering parameters such as the scattering length and the effective range are very important physical quantities that are used to describe a variety of systems. Among these are nuclear reactions of relevance to nucleosynthesis \cite{Rolfs} and
cold dilute atomic gases \cite{Pethik}. To find the low energy scattering behaviour for a given potential, the generic approach is to solve the stationary Schr\"odinger equation for a range of scattering momenta and to extract the scattering phase shifts \cite{Flugge}. In contrast, our formalism allows the computation of the scattering length and effective range directly from the potential.

To reformulate quantum scattering theory, we use methods from classical mechanics. A "Hamiltonian" function
expressed in terms of the coefficients of the Riccati-Bessel and Riccati-Neumann function expansion of the solution
of the Schr\"odinger equation was found amenable to formal canonical manipulations. This Hamiltonian, of a quadratic
form, can be rendered constant with an appropriate canonical transformation. From this transformation we obtain a first order, nonlinear differential equation for the scattering problem that generalises Calogero's equation \cite{Calogero}. We demonstrate how this equation can be used to solve the scattering problem for low momenta and several partial waves. Our examples range from nuclear scattering to atomic physics.

In the latter, the basic input into the atom-atom effective interaction is the scattering length $a$, defined as the scattering amplitude evaluated at zero energy $a = - f(E=0)$, where $f$ is determined from the $l=0$ phase shift, $\delta_{0}(k)$.

The effective range expansion for $D_{l}(k)=\frac{\tan(\delta_{l}(k))}{k^{2l+1}}$ is \cite{Teichmann},
\begin{equation}
\frac{1}{D_{l}(k)} = -\frac{1}{a_{l}} + \frac{1}{2} r_{l} k^{2} - P_{l} r_{l}^{3} k^{4}.
\label{effectiverangeexpansion}
\end{equation}
Our formalism allows the calculation of $a_{l}$, $ r_{l}$ and $P_{l}$ directly from the potential, using first order, non-linear, differential equations, and may allow the development of new approximation schemes for the calculation of these low energy parameters in cases where the usual procedure of solving the second order equation (the Schr\"odinger equation) may be cumbersome.\\

This paper is organized as follows: In \sref{canonical} we present our canonical formulation of potential scattering. In \sref{slfunction} we derive the Calogero-type equation for the scattering length function and obtain first order non-linear differential equations for the scattering length, effective range and shape parameter. In \sref{coldatoms} we apply the theory to the calculation of the low-energy parameters of typical atomic scattering systems. One further application to typical short range potential nuclear scattering is presented in \sref{nuclear}. Finally, in \sref{conclusion} several concluding remarks are made. Appendices contain the essentials of potential scattering and generalisations of our work to charged particle scattering.

\bigskip

\section{Canonical Formulation}
\label{canonical}

In quantum scattering from a central potential $V(r) = \frac{\hbar^2}{2m}U(r)$ one has to solve the radial Schr\"odinger equation \cite{Newton, CH},
\begin{equation}
\big[-\frac{d^{2}}{dr^{2}} + U(r)+ \frac{l(l+1)}{r^{2}}\big]\phi_{l}(k,r) = k^{2} \phi_{l}(k,r),
\label{radialSE}
\end{equation}
where $l$ is the angular momentum quantum number and $k$ the scattering momentum. It is known that the solution can be parametrised as
\begin{equation}
\phi_{l}(k,r) = u_{l}(kr) q_{l}(k,r) + v_{l}(kr) p_{l}(k,r),
\end{equation}
where $u_{l}(kr)$ and $v_{l}(kr)$ are the Riccati-Bessel and Riccati-Neumann functions.

Following \cite{Hara,Hussein}, we construct a classical "Hamiltonian" function,
\begin{equation}
H(q,p) = \frac{U(r)}{2k} \big( u_{l} q_{l} + v_{l} p_{l} \big)^2.
\end{equation}
Through a straightforward computation of $\frac{dq_{l}}{dr}$ and $\frac{dp_{l}}{dr}$ from Eq.~(\ref{radialSE}) and Eq.~(\ref{freeradialSE}), (\ref{wronskian})-(\ref{plkr}) in the Appendix, we find, 
\begin{equation}
\frac{dq_{l}}{dr} = \frac{\partial H}{\partial p_{l}},
\end{equation}
\begin{equation}
\frac{dp_{l}}{dr} = - \frac{\partial H}{\partial q_{l}}.
\end{equation}

\bigskip

It is therefore evident that we may regard $q_{l}$ and $p_{l}$ as the coordinate and conjugate momentum respectively, with $r$ acting as the 
"time" variable in a "classical" dynamics governed by the Hamiltonian $H(q,p)$ above \cite{Goldstein}. It should be realized that in this fashion we treat each partial wave independently as there is no coupling between waves for different $l$ due to the spherical symmetry of $V(r)$. Thus the scattering problem
is reduced to a many- (strictly speaking infinite) non-interacting classical fields problem.

Any function $F_{l}(q,p;r)$ of $q$, $p$, and $r$ will evolve in "time", $r$, as
\begin{equation}
\frac{dF_{l}}{dr} = [F_{l}, H] + \frac{\partial F_{l}}{\partial r},
\end{equation}
where $[F_{l},H]$ is the Poisson bracket.

In order to identify useful functions for the scattering problem, such as the phase shift, we perform a canonical transformation
to a new coordinate, $Q_{l}$ and a new momentum $P_{l}$, using the following generating function,
\begin{equation}
F_{2} ( q_{l}, P_{l},r) = \frac{1}{2} A_{l} (r) q_{l}^{2}  + B_{l}(r) q_{l}P_{l},
\end{equation}
where the functions $A_{l}(r)$ and $B_{l}(r)$ are arbitrary to be determined by imposing a condition on the new dynamics, and  
\begin{equation}
Q_{l} = \frac{\partial F_{2}}{\partial P_{l}} = q_{l} B_{l}(r),
\end{equation}
and
\begin{equation}
p_{l} = \frac{\partial F_{2}}{\partial q_{l}} = A_{l}(r) q_{l} + B_{l}(r) P_{l}.
\end{equation}
The new Hamiltonian function $K(Q_{l},P_{l};r)$ is
\begin{equation}
K (Q_{l}, P_{l};r) = H(q_{l}, p_{l};r) +\frac{\partial F_{2}(q_{l}, P_{l};r)}{\partial r}.
\end{equation}
It is now a simple matter to obtain the explicit form of the new Hamiltonian, $K$, viz
\begin{eqnarray}
K(Q_{l},P_{l};r) = \frac{Q_{l}^{2}}{2B_{l}^2} \big[ \frac{U(r)}{k}( u_{l} + v_{l}A_{l})^2 + \frac{dA_{l}}{dr}\big]\nonumber\\
+\frac{Q_{l}P_{l}}{B_{l}} \big[ \frac{U(r)}{k} v_{l}(u_{l} + v_{l} A_{l}) B_{l} + \frac{d B_{l}}{d r}\big]\nonumber\\
+\frac{U(r)}{2 k}v_{l}^2 B_{l}^2 P_{l}^2
\end{eqnarray}
The goal is now to render the new coordinate $Q_{l}$ cyclic. To this end, we impose the condition that the functions $A_{l}(r)$ and $B_{l}(r)$ satisfy the first-order, non-linear differential equations,
\begin{equation}
\frac{d A_{l}(r)}{d r} + \frac{U(r)}{ k} \big( u_{l}(r) + v_{l}(r) A_{l}(r) \big)^{2} = 0,
\label{dAdreqn}
\end{equation}
and
\begin{equation}
\frac{d B_{l}(r)}{d r} + \frac{U(r)}{k} \big( u_{l}(r) + v_{l}(r) A_{l}(r) \big) B_{l}(r) =0.
\end{equation}
To solve these, we need to specify a boundary condition. We choose the values of the functions
$A_{l} (0) =0$ and $B_{l}(0) = 1$, which guarantee that the new coordinate $Q_{l}$ and momentum $P_{l}$ have the same limiting values at $r=0$
as the original ones, $q_{l}$ and $p_{l}$, namely $Q_{l}(r)\rightarrow 1$ and $P_{l}(r)\rightarrow 0$ as $r\rightarrow 0$. 

With this choice of the coefficients $A_{l}(r)$ and $B_{l}(r)$, the new Hamiltonian function $K(Q_{l},P_{l};r)$ is
\begin{equation}
K(Q_{l}, P_{l}; r) = \frac{U(r)}{2k} v_{l}^{2}(r) B_{l}^{2}(r) P_{l}^{2}
\end{equation}
and the new equations of motion,
\begin{equation}
\frac{dP_{l}}{dr} = - \frac{\partial K}{\partial Q_{l}} = 0
\end{equation}
and
\begin{equation}
\frac{dQ_{l}}{dr} = \frac{\partial K}{\partial P_{l}} = \frac{ U(r)}{2k} v_{l}^{2}(r)B_{l}^{2} P_{l}.
\end{equation}
We choose the solutions, $P_{l}(r) = 0$ and $Q_{l}(r) = 1$, which are consistent with the new equations of motion and the boundary conditions.
With this choice, we have the solution of the original problem,
\begin{equation}
q_{l}(r) =\frac{1}{B_{l}(r)}
\end{equation}
and
\begin{equation}
p_{l}(r) = \frac{A_{l}(r)}{B_{l}(r)}.
\end{equation}
By elimination of $B_{l}(r)$ from the above equations, we obtain for the function $A_{l}(r)$
the following
\begin{equation}
A_{l}(r) = \frac{p_{l}(r)}{q_{l}(r)},
\end{equation}
which can be identified as the phase shift function $\tan{\delta_{l}(r)}$, introduced by Calogero \cite{Calogero}.

For charged particle scattering, the 
free solutions $u_{l}(kr)$, $v_{l}(kr)$ are replaced by the scaled Coulomb wave functions, defined in \aref{charged}. The equation for the tangent function $A_{l}(r)$ is given by a similar
equation as \eref{dAdreqn}, with the replacement of $u_{l}(kr)$ by the regular scaled Coulomb wave function, $\mathcal{F}_{l}(kr)$ and $v_{l}(kr)$ by the irregular
scaled Coulomb wave function, $\mathcal{G}_{l}(kr)$. A detailed discussion of the charged particle Calogero equation of the scattering length function is beyond the scope of this paper, but a brief preview is given in \aref{charged}.

\section{Scattering Length Function}
\label{slfunction}

It is convenient to introduce the scattering length function $a(k,r)$ defined as the function,
\begin{equation}
a_{l}(k,r) = - \frac{A_{l}(r)}{k^{(2l+1)}},
\end{equation}
which satisfies the Calogero-type equation,
\begin{equation}
\frac{da_{l}(k,r)}{dr} - \frac{U(r)}{k^{2l+2}}\big(u_{l}(kr) - k^{(2l+1)} v_{l}(kr) a_{l}(k,r)\big)^{2} = 0.
\label{maineqn_anyk_anyl}
\end{equation}
Note that $a_{l}(k;\infty)$ has the dimension of $L^{2l+1}$.

The above equation has to be solved with the boundary condition $a_{l}(k,0) = 0$. The scattering length itself is, by definition, $a_{l}(0,\infty)$.
For $l=0$ the equation becomes,
\begin{equation}
\frac{da(k,r)}{dr} - \frac{U(r)}{k^2} \big( \sin(kr) - k \cos(kr) a(k,r) \big)^{2} = 0
\label{maineqn_anyk_zerol}
\end{equation}
Further specifying $k=0$ we have
\begin{equation}
\frac{da(0,r)}{dr} - U(r)\big[r - a(0,r)\big]^{2} = 0.
\label{maineqn_zerok_zerol}
\end{equation}
A final special case of interest is $k=0$ and integer $l$. Then the asymptotic behaviour of the Bessel functions (see Eqs.~\ref{usmall}-\ref{vsmall}) gives:
\begin{equation}
\frac{da_{l}(0,r)}{dr} - U(r)\left[\frac{r^{l+1}}{(2l+1)!!} - \frac{(2l-1)!!}{r^l}a_{l}(0,r)\right]^{2} = 0.
\label{maineqn_zerok_anyl}
\end{equation}
\erefs{maineqn_anyk_anyl}{maineqn_zerok_anyl} are already sufficient to extract the low energy scattering behaviour that defines \eref{effectiverangeexpansion} as we shall demonstrate in the next section. However they can also be used to obtain equations that determine the parameters of the effective range expansion directly. This can be done by expanding \eref{maineqn_anyk_anyl} in $k$, equating coefficients and using the identification $D_{l}(k)\rightarrow- a_{l}(k,r)$. The resulting Calogero equations for $r_{0}(r)$ and $p_{0}(r) = P_{0}(k,r)r_{0}(k,r)^3$ are
\begin{equation}
\frac{dr_0(r)}{dr} + 2rU(r)r_{0}(r)(\frac{r}{a_{0}(r)}-1) + 2r^{2} U(r)[\frac{r}{a_{0}(r)}-1][\frac{r}{3a_{0}(r)}-1] = 0,
\label{effrangeeqn}
\end{equation}
\begin{equation}
\frac{dp_{0}(r)}{dr} + 2rU(r)[\frac{r}{a(r)}-1]p_{0}(r) + F(r) = 0
\label{shapeparameqn}
\end{equation}
where the function $F(r)$ is given by,
\begin{equation}
F(r) = \frac{r^2U(r)}{12}(2r -r_{0}(r))(2r -3r_{0}(r))+ \frac{r^2U(r)}{45a_{0}^2(r)}[2r^4 +3a_{0}(r)r^2(5r_{0}(r)-4 r)]
\end{equation}
The above two differential equations can be integrated once the $l=0$ scattering length function, $a_{0}(r)$ is known. Numerically, \erefs{maineqn_zerok_zerol}{shapeparameqn} could be solved as coupled system. Notice that $r_{0}(r=0)= 0$ and $p_{0}(r =0) = 0$, owing to the boundary condition obeyed by the scattering length function, $a_{l}(k, r =0)= 0$. 

The scattering length equation for any partial wave can be written down, with the aid of \eref{maineqn_anyk_anyl} and \erefs{usmall}{vsmall} of the Appendix  
\begin{equation}
\frac{da_{l}(0,r)}{dr} - U(r)\big[\frac{r^{l+1}}{(2l+1)!!} - \frac{(2l-1)!!}{r^l} a_{l}(0,r)\big]^2 = 0,
\end{equation}
The corresponding $l \neq 0$ version of the Calogero equations for the effective range function, $r_{l}(r)$, and shape parameter function, $p_{l}(r)$, can be easily worked out using the small $k$ form of the Riccati-Bessel and Riccati-Neumann functions, given in \aref{quantumscattering}. 

\section{Cold atom scattering}
\label{coldatoms}

To show how our formalism simplifies typical calculations, we now use it to determine low energy scattering parameters for some exemplary potentials and compare our results with the literature. We shall mostly consider scattering of Cs atoms interacting with the potential \cite{Flambaum}
\begin{eqnarray}
V(r)=\frac{1}{2}\beta r^{\lambda}e^{-\eta r}-\left(\frac{C_{6}}{r^{6}} + \frac{C_{8}}{r^{8}} + \frac{C_{10}}{r^{10}} \right)f_{c}(r),
\\
f_{c}(r)=\theta(r-r_{c}) + \theta(r_{c}-r)e^{-(r_{c}/r-1)^{2}},
\end{eqnarray}
where in atomic units $\beta=1.6\times 10^{-3}$, $\lambda=5.53$, $\eta=1.072$, $C_{6}=7020$, $C_{8}=1.1\times 10^6$, $C_{10}=1.7\times 10^8$, $r_c=23.1654$ and the reduced mass of a cesium pair is $m=1.211 \times 10^5$.

\subsection{S-wave scattering length}

The s-wave scattering length can be determined according to the Calogero equation~(\ref{maineqn_zerok_zerol}). It has been recently used to assess corrections to calculations that rely on
using the usual Schr\"odinger equation with long range interactions \cite{Ouer}. Here we first summarize their calculations and then extend them to other partial waves and to determining the effective range and shape parameters.

Integrating this equation is numerically non-trivial, owing to poles in $a(0,r)$ that arise from bound states in the potential. This problem can be overcome by using a symplectic integrator or a variable transformation that maps the spatial domain and the range of function values onto a compact interval \cite{Ouer}. For the latter one defines
$\tan[\theta(r)]=a(0,r)$ and $r=\tan(\phi)$ and then solves a nonlinear equation for $\theta(\phi)$. Since \eref{maineqn_zerok_zerol} is singular at $r=0$ the initial condition is $a(\epsilon)=0$, with $\epsilon$ chosen sufficiently small. The determination of the Cs-Cs scattering length in this manner has been done in \rref{Ouer}, which also shows $a(0,r)$ as a function of $r$ to visualize the above mentioned pole structure. The scattering length following from this potential, $a_{s}=68.21$ is reproduced precisely by \eref{maineqn_zerok_zerol}.

\subsection{S-wave effective range expansion}
\label{expansion}

Our formalism allows us to use the transformation of the preceding section to determine quantities beyond the scattering length. For the effective range, we present two methods: the direct integration of \eref{effrangeeqn} and the integration of \eref{maineqn_anyk_zerol} for a range of $k$ and subsequent fitting of  \eref{effectiverangeexpansion}.

\subsubsection{Effective range equation}
\begin{figure}
\centering
\epsfig{file={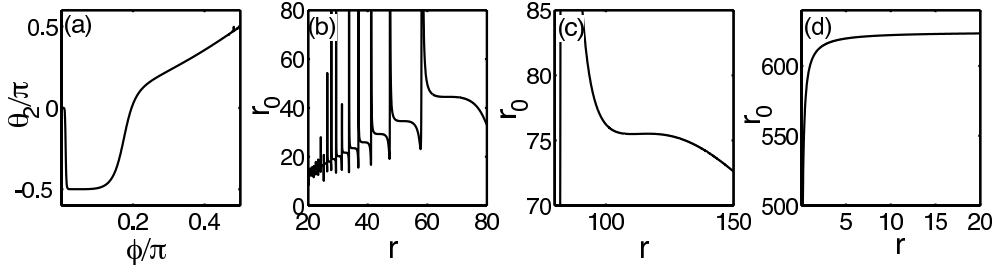},width=\columnwidth} 
\caption{(a) Function $\theta_{2}(\phi)$ obtained from \eref{directeffrange}, (b-d) Reconstructed $r_{0}(r)$ for different ranges of $r$. The pole structure in (b) arises from bound states in the potential. The asymptotic value reached in (d) gives the physical effective range $r_{0}=623.361$.
\label{effrangesolution}}
\end{figure}
We use the substitutions $\tan[\theta_{1}(r)]=a(0,r)$, $\tan[\theta_{2}(r)]=r_{0}(r)$ and $r=\tan(\phi)$ in \erefs{maineqn_zerok_zerol}{effrangeeqn}. The resulting system of two coupled equations is:
\begin{eqnarray}
\frac{d \theta_{1}(\phi)}{d \phi}-
\sec^4{[\phi]}\sin^2{[\theta_{1}(\phi)-\phi]}U[\tan(\phi)]=0,
\nonumber
\\
\frac{d \theta_{2}(\phi)}{d \phi}-
\sec^2{[\phi]}\cos^2{[\theta_{2}(\phi)]}U[\tan(\phi)]
\bigg[
 2 \tan[ \theta_{2}(\phi)] \tan(\phi)
 \left(1-\frac{\tan(\phi)}{\tan[\theta_{1}(\phi)]}\right)
\nonumber
\\
 -2\tan(\phi)^2 + \frac{8}{3}\frac{\tan(\phi)^3}{\tan[\theta_{1}(\phi)]} - \frac{2}{3}\frac{\tan(\phi)^4}{\tan[\theta_{1}(\phi)]^2}
 \bigg]=0,
\label{directeffrange}
\end{eqnarray}
Again we use the initial conditions $a(0,\epsilon_{r})=\epsilon_{a}$ and $r_{0}(\epsilon_{r})=0$, with $\epsilon_{a}$, $\epsilon_{r}$ small, to avoid an initial singularity. The solution is shown in \fref{effrangesolution}, similar plots showing the scattering length can be found in \rref{Ouer}.

\subsubsection{Low $k$ expansion}
\begin{figure}
\centering
\epsfig{file={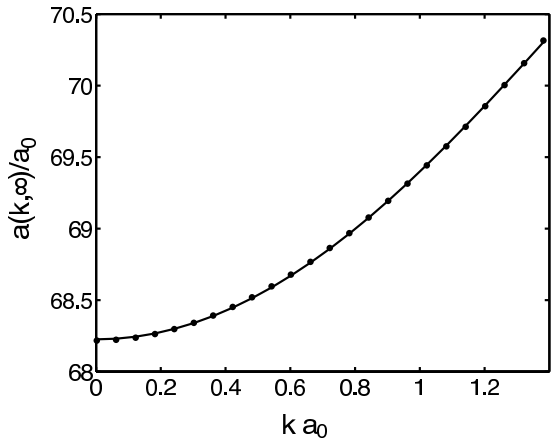},width=0.5\columnwidth} 
\caption{Asymptotic s-wave scattering length function ($\bullet$) for several values of wave number $k$. The solid line is a least squares fit of the function $D_{0}(k)$ (see \eref{effectiverangeexpansion}), from which we can extract the scattering length $a_{s}=68.23\pm0.01$ $a_{0}$, the effective range $r_{s}=541.69\pm6$ $a_{0}$ and the shape parameter $P_{s}=0.144\pm0.006$. The uncertainties reflect $95\%$ confidence bounds of the fit.
\label{lowksolution}}
\end{figure}

Alternatively, in \eref{maineqn_anyk_zerol} we define $\tan[\theta(k,r)]=a(k,r)$ and $r=\tan(\phi)$. This yields
\begin{eqnarray}
\frac{d \theta(k, \phi)}{d \phi}-
 \sec^2{[\phi]}\cos^2{[\theta(k, \phi)]} \frac{V[\tan(\phi)]}{E_{k}} \big( \sin[k\tan(\phi)] - k \cos[k\tan(\phi)] \tan[\theta(k, \phi)] \big)^{2}.
\label{maineqn_anyk_zerol_transf}
\end{eqnarray}
We now solve \eref{maineqn_anyk_zerol_transf} for several values of $k$, see \fref{lowksolution}. Using the effective range expansion as in \eref{effectiverangeexpansion} we can then extract the low energy scattering parameters from a fit as shown. The values for $a_{s}$ and $r_{s}$ agree roughly those of \rref{Flambaum} (which are $a_{s}=68.22$ and $r_{s}=624.55$). 
For the fit we varied the range of values of $k$ under consideration until the smallest errors were reported by the fitting routine.
The directly calculated effective range agrees better with \rref{Flambaum} than that obtained from the fit. However the fitting procedure is able to also give an indication of the shape parameter, whereas the direct determination of it, using \eref{shapeparameqn}, was plagued by numerical instabilities.

\newpage

\subsection{Higher partial waves}

The scattering lengths for higher partial waves (p,d,f,...) can in principle be obtained from the simple \eref{maineqn_zerok_anyl}. For the potential given at the beginning of this section we find $a_{p}=-4.38\times 10^{5}$ $a_{0}^3$. Higher partial waves for the atom-atom potential cannot be treated with our method, since the centrifugal terms in \eref{maineqn_zerok_anyl} combine with the long range tail of the potential to produce a constant slope of the scattering length function. This shortcoming
is absent in the case of short range interaction, such as the one encountered in neutron-nucleus scattering, as we discuss in the next section.
As additional check of our theory we reproduced the s-wave and p-wave scattering length for $^3$He and $^4$He given in \cite{Gutierrez}, which relate to the HFDHE2 potential of \cite{Azis}. This comparison is shown in \tref{pwavecomparison}.

\begin{table}
\begin{center}
\begin{tabular}{|c|cc|}
\cline{1-3}
\Big. 
parameter      & present study      & \rref{Gutierrez}  \\
\cline{1-3}
$^3$He  $a_{s}$    & $-7.03$ $\angstrom$  & $-6.97$ $\angstrom$   \\
$^3$He  $a_{p}$    & $-26.15$  $\angstrom^3$  & $-26.09$ $\angstrom^3$  \\
$^4$He  $a_{s}$    & $97.21$ $\angstrom$    & $125.08$ $\angstrom$   \\
$^4$He  $a_{p}$    & $-42.67$ $\angstrom^3$   & $-42.12$ $\angstrom^3$    \\
\cline{1-3}
\end{tabular}
\end{center}
\caption{S-wave and p-wave scattering length for the two helium isotopes.
\label{pwavecomparison}}
\end{table}
%
\section{Nuclear scattering}
\label{nuclear}

For potentials that drop off faster than any polynomial at large $r$ all the partial waves can be obtained.
To demonstrate this, we apply our method to calculate the low energy scattering parameters
for the $n$ + $^{12}$C system, where the interaction is short-range and contains a spin-orbit part \cite{Baye}. In this latter case, owing to the spin $s$ of the neutron, one must specify the total angular momentum, ${\bf j} = {\bf l} + {\bf s}$, which can take either one of two values $j = l $+1/2 or $j =l$-1/2. The scattering length and the other low-energy parameters acquire two labels $jl$. The interaction itself is split into two,
\begin{equation}
V(r) = V_{0}(r) + V_{so}(r) {\bf l} \cdot {\bf s}.
\end{equation}
For j = l+1/2, one has
\begin{equation}
V_{j=l+1/2}(r) = V_{0}(r) + \frac{l}{2}V_{so}(r)
\end{equation}
and for j = l - 1/2, one has
\begin{equation}
V_{j=l-1/2}(r) = V_{0}(r) - \frac{l+1}{2} V_{so}(r).
\end{equation}

In the above equations the spherical potential is denoted by $V_{0}(r)$ and is invariably taken to have a Fermi-type dependence on $r$, the so-called the Woods-Saxon shape
\begin{equation}
V_{0}(r) = - \frac{V_{0}}{1 + \exp{(\frac{r - R}{d})}},
\end{equation}
while the spin-orbit potential $V_{so}(r)$ is given by

\begin{equation}
V_{so}(r) = \frac{5.5}{1 + \exp{(\frac{r - R}{d})}}.
\end{equation}

As in \sref{expansion}, we extract the scattering length and effective range from a fit to the low momentum behaviour of $a_{l,s}(k,\infty)$. A comparison of our results with those of \rref{Baye} is shown in \tref{nuclearcomparison}. As in \cite{Baye} we choose $R = 2.86$ fm, and $d = 0.65$ fm, while the strength of the central potential $V_{0}$ is adjusted for each n-$^{12}$C bound-state energy.
\begin{table}
\begin{center}
\begin{tabular}{|c|cccccc|}
\cline{1-7}
\Big. 
$l,s$      & $V_{0}$ & $\underline{a_{l,s}}$      & $a_{l,s}$,  \rref{Baye}  & $\underline{r_{l,s}}$      & $r_{l,s}$,  \rref{Baye} & $\underline{p_{l,s}}$\\
\cline{1-7}
$s1/2$ & $57.6$  & $6.51$ & $6.43$  & $3.58$ & $3.56$  & $-0.055$  \\
$p3/2$ & $45.1$  & $9.16$ & $8.85$  & $-1.68$ & $-1.71$  & $0.038$    \\
$p1/2$ & $45.1$  & $23.21$ & $22.75$  & $-1.15$ & $-1.16$  & $0.26$    \\
$d5/2$ & $56.15$  & $179.6$ & $159.9$  & $-0.32$ & $-0.32$  & $-28.65$    \\
$d3/2$ & $56.15$  & $-56.1$ & $-57.2$  & $ -0.061$ & $-0.065$  & $-2864$    \\
\cline{1-7}
\end{tabular}
\end{center}
\caption{Scattering length (in fm$^{2l+1}$) and effective range (in fm$^{-2l+1}$) for different angular momentum channels of a $n$ + $^{12}$C collision \cite{Baye}. Underlines indicate the results from the present study. The well depth of the channels, $V_{0}$ (in MeV), is also shown.
\label{nuclearcomparison}}
\end{table}

\section{Conclusion}
\label{conclusion}

In this paper we have explored an analogy between quantum potential scattering and the classical dynamics of a conservative system of infinite fields. With the aid of an appropriate canonical transformation of a classical Hamiltonian, we were able to derive the Calogero equation for the tangent of the phase shift function. This allowed us to obtain a first order non-linear differential equation for the so-called scattering length function, whose asymptotic value for large
separations supplies the well known low-energy expansion in terms of the scattering length, effective range and shape parameter for any partial wave. We have applied our theory to obtain these parameters for typical atom-atom systems (long-range interaction) and neutron-nucleus systems (short-range interaction). We reached very good agreements with results obtained through conventional methods like a direct solution of the scattering Schr\"odinger equation. 

\appendix
\section{Essentials of Quantum Potential Scattering Theory}
\label{quantumscattering}

In this appendix we supply the essentials of quantum scattering theory as required in \sref{canonical}. We consider the scattering of a particle from a spherically symmetric potential. The extension to the scattering of a spin-1/2 particle can be easily
formulated by adding a spin-orbit interaction term to the potential.
The regular solution of the radial Schr\"odinger equation describing the scattering by a spherical potential, $V(r) = \frac{\hbar^2}{2m}U(r)$ \cite{Newton, CH},
\begin{equation}
\big[-\frac{d^{2}}{dr^{2}} + U(r)+ \frac{l(l+1)}{r^{2}}\big]\phi_{l}(k,r) = k^{2} \phi_{l}(k,r)
\label{radialSE_App}
\end{equation}
can be written as,
\begin{equation}
\phi_{l}(k,r) = u_{l}(kr) q_{l}(k,r) + v_{l}(kr) p_{l}(k,r),
\end{equation}
where $u_{l}(kr)$ and $v_{l}(kr)$ are the Riccati-Bessel and Riccati-Neumann functions, defined in terms of the usual spherical Bessel, $j_{l}(kr)$, and spherical Neumann functions, $n_{l}(kr)$, respectively,
\begin{equation}
u_{l}(kr) = krj_{l}(kr),
\end{equation}
\begin{equation}
v_{l}(kr) = krn_{l}(kr).
\end{equation}
The functions, $u_{l}(kr)$ and $ v_{l}(kr)$ are the regular and irregular solutions of the free radial Schr\"odinger equation
\begin{equation}
\big[-\frac{d^{2}}{dr^{2}} + \frac{l(l+1)}{r^{2}}\big] \omega_{l}(kr) = k^{2} \omega_{l}(kr),
\label{freeradialSE}
\end{equation}
where $\omega_{l}(kr)$ is $u_{l}(r)$ or $ v_{l}(r)$.
Clearly the coefficients $q_{l}(k,r)$ and $p_{l}(k,r)$ contain all information about the scattering.

\bigskip

Since $\phi_{l}(k,r)\rightarrow_{r \rightarrow 0}u_{l}(kr)$, we have the following  boundary conditions satisfied by $q_{l}(k,r) $ and $p_{l}(k,r)$
\begin{equation}
q_{l}(k,r)\rightarrow_{r \rightarrow 0} 1,
\end{equation}
\begin{equation}
p_{l}(k,r)\rightarrow_{r \rightarrow 0} 0.
\end{equation}
Since the functions $u_{l}(k,r)$ and $v_{l}(k,r)$ are linearly independent in the sense that the Wronskian,
\begin{equation}
W[v,u] = v u^{\prime} - v^{\prime} u = k,
\label{wronskian}
\end{equation}
it follows that $q_{l}(k,r)$ and $p_{l}(k,r)$ may be expressed as
\begin{equation}
q_{l}(k,r) = \frac{1}{k}W[v_{l},\phi_{l}],
\label{qlkr}
\end{equation}
\begin{equation}
p_{l}(k,r) = -\frac{1}{k}W[u_{l},\phi_{l}].
\label{plkr}
\end{equation}

The usual method of obtaining the scattering observables, is to integrate the radial equation and adjust the asymptotic form to the following
\begin{equation}
\phi_{l}(k, r) \rightarrow \sin(kr - l\frac{\pi}{2} + \delta_{l}(k)),
\end{equation}
which allows the extraction of the phase shift function $\delta_{l}(k)$. All observables can be written in terms of $\delta_{l}(k)$. For example, the scattering amplitude $f(k,\theta)$ is just
\begin{equation}
f(k, \theta) = \frac{1}{2ik}\sum_{l=0}^{\infty} (2l+1)\big(1 - \exp{[2i\delta_{l}(k)]}\big)P_{l}(\cos{\theta}),
\end{equation}
with the low-energy limit ( only $l=0$ )
\begin{equation}
f(k,0) = \frac{1}{k}\exp(i\delta_{0}(k))\sin(\delta_{0}(k)) = \frac{1}{k \cot{\delta_{0}(k) - i k}}.
\end{equation}
where the function $-\frac{\tan{\delta_{0}(k)}}{k}$, in the limit of zero energy, is identified with the scattering length.

In the following we enumerate some useful formulae for the Riccati-Bessel and Riccati-Neumann functions used in the last section.

The Riccati-Bessel function is given by
\begin{equation}
u_{l}(\rho)= -(-\rho)^{l+1}\big(\frac{1}{\rho}\frac{d}{d\rho}\big)^{l}j_{0}(\rho),
\end{equation}
where the spherical Bessel function $j_{0}(\rho) = \frac{\sin{\rho}}{\rho}$, and $\rho = kr$.
Similar relation holds for the Riccati-Neumann function
\begin{equation}
v_{l}(\rho) = -(-\rho)^{l+1}\big(\frac{1}{\rho}\frac{d}{d\rho}\big)^{l} n_{0}(\rho),
\end{equation}
where the spherical Neumann function $n_{0}(\rho)= \frac{\cos{\rho}}{\rho}$.\\
The recursion relation follows
\begin{equation}
\omega_{l}(\rho) = (2l +1)\frac{\omega_{l+1}(\rho)}{\rho} - \omega_{l-1}(\rho),
\end{equation}
where $\omega_{l}(\rho)$ stands for $u_{l}(\rho)$ or $v_{l}(\rho)$.
For the $l=1$  partial wave, we have
\begin{equation}
u_{1}(\rho) = \frac{\sin{\rho}}{\rho} - \cos{\rho}
\end{equation}
and
\begin{equation}
v_{1}(\rho) = \frac{\cos{\rho}}{\rho} + \sin{\rho}.
\end{equation}
For higher values of $l$, these functions are easily obtained from the recursion relation above.\\

\bigskip

Finally, the behaviour of these functions for small values of the argument is
\begin{equation}
u_{l}(\rho \rightarrow 0) = \frac{\rho^{l+1}}{(2l+1)!!}
\label{usmall}
\end{equation}
and
\begin{equation}
v_{l}(\rho \rightarrow 0) = \frac{(2l-1)!!}{\rho^{l}}.
\label{vsmall}
\end{equation}

\section{Charged Particle Scattering}
\label{charged}

When the scattering particles are charged, the $\frac{1}{r}$ Coulomb interaction has to be added to the potential. The free solutions are now Coulomb wave functions and special care must be taken when performing the low-energy expansion to ensure convergent results for the scattering length, effective range and shape parameters. The appropriate low-energy expansion was developed in \cite{Teichmann}, and the quantity $D_{l}(E)$, becomes in this case
\begin{equation}
D_{l}^{c}(E) = \frac{\pi}{2}\exp{(2\pi\eta)}\tan{\delta_{l}(E)}
\end{equation}
where $\eta$ is the Sommerfeld parameter defined in terms of the charges of the two particles $Z_1$ and $Z_2$, by $\eta = \frac{Z_1 Z_2 e^2}{\hbar v}$, with $v$ being the asymptotic relative velocity. The low-energy expansion is then given by
\begin{equation}
2\frac{\omega_{l}(E)}{(l!)^2 a_{N}^{2l+1}}\big[\frac{2}{D_{l}^{c}(E)} + h(\eta) \big] = -\frac{1}{a_{l}} + \frac{1}{2} r_{l} k^{2} - P_{l} r_{l}^{3} k^{4} + O(k^6).
\end{equation}
In the above, the functions $\omega_{l}(E)$ and $h(\eta)$ are, respectively
\begin{equation}
\omega_{l}(E) = \prod_{n=1}^{l}\big(1 + \frac{n^2}{\eta^2} \big),
\end{equation}
\begin{equation}
h(\eta) = \frac{1}{12\eta^2} + \frac{1}{120\eta^4} + O\big(\frac{1}{\eta^6}\big),
\end{equation}
and $a_{N}$ is the so-called nuclear Bohr radius, $a_{N} = \frac{\hbar^2}{\mu Z_1 Z_2 e^2}$, with $\mu$ being the reduced mass. 
The above equation does indeed give convergent results as $\eta$ goes to $\infty$ in the zero-$E$ limit. Further, the equation reduces to that for neutral particles upon setting $\eta =0$. 

The corresponding Calogero equation then follows by use of the Wronskian of the scaled Coulomb wave functions
\begin{equation}
W[\mathcal{G}_{l}(kr),\mathcal{F}_{l}(kr)] = \frac{\pi}{2},
\end{equation}
where $\mathcal{F}_{l}(kr)$ and $\mathcal{G}_{l}(kr)$ are related to the regular, $F_{l}(kr)$, and irregular, $G_{l}(kr)$, Coulomb wave functions \cite{AS}
\begin{equation}
\mathcal{F}_{l}(kr) = k^{-1/2}\exp{(\pi\eta)} F_{l}(kr),
\end{equation}
and
\begin{equation}
\mathcal{G}_{l}(kr) = \frac{\pi}{2} k^{-1/2}\exp{(-\pi\eta)}G_{l}(kr).
\end{equation}
The "Hamiltonian" function is 
\begin{equation}
H(q_{l}, p_{l}, r) = \frac{U(r)}{\pi} (\mathcal{F}_{l}q_{l} + \mathcal{G}_{l}p_{l})^2.
\end{equation}
The Calogero equation for the tangent function can be derived following the same procedure as the one used for the neutral particle scattering
\begin{equation}
\frac{da^{c}_{l}(k,r)}{dr} - \frac{\pi}{2}\exp{[-2\pi\eta]} U(r) \big[ \mathcal{F}_{l}(kr) - \frac{\pi}{2}\exp{[-2\pi\eta]}\mathcal{G}_{l}(kr) a^{c}_{l}(k,r)\big]^2 = 0,
\end{equation}
where $a^{c}_{l}(k,\infty)$ is just the Coulomb modified tangent function, $D_{l}^{c}(E) = \frac{\pi}{2}\exp{(2\pi\eta)}\tan{\delta_{l}(E)}$. Clearly, once
this function is calculated, the low energy parameters can be obtained from the expression $2\frac{\omega_{l}(E)}{(l!)^2 a_{N}^{2l+1}}\big[\frac{2}{D_{l}^{c}(E)} + h(\eta) \big] = -\frac{1}{a_{l}} + \frac{1}{2} r_{l} k^{2} - P_{l} r_{l}^{3} k^{4} + O(k^6)$, above.

\end{document}